\documentclass[aps,pre,showpacs,twocolumn]{revtex4}
\usepackage{graphics}
\usepackage{epsfig}

\begin{document}

\title{Force distributions in a triangular lattice of rigid bars}

\author{Brian P.~Tighe}
\affiliation{Department of Physics and Center for Nonlinear and Complex Systems, \\
         Duke University, Durham, NC 27708}
\author{Joshua E.~S.~Socolar}
\affiliation{Department of Physics and Center for Nonlinear and Complex Systems, \\
         Duke University, Durham, NC 27708}
\author{David G.~Schaeffer}
\affiliation{Department of Mathematics and Center for Nonlinear and Complex Systems, \\
         Duke University, Durham, NC 27708}
\author{W.~Garrett Mitchener}
\affiliation{Department of Mathematics and Center for Nonlinear and Complex Systems, \\
         Duke University, Durham, NC 27708}
 \author{Mark L.~Huber}
\affiliation{Department of Mathematics and Center for Nonlinear and Complex Systems, \\
         Duke University, Durham, NC 27708}

\date{\today}

\begin{abstract} 
  We study the uniformly weighted ensemble of force balanced
  configurations on a triangular network of nontensile contact forces.
  For periodic boundary conditions corresponding to isotropic
  compressive stress, we find that the probability distribution for single-contact 
  forces decays faster than exponentially.  This
  super-exponential decay persists in lattices diluted to the rigidity
  percolation threshold.  On the other hand, for anisotropic imposed
  stresses, a broader tail emerges in the force distribution, becoming
  a pure exponential in the limit of infinite lattice size and
  infinitely strong anisotropy.
\end{abstract}

\maketitle

\section{Introduction}
Materials composed of hard cohesionless grains, such as dry sand,
exhibit many remarkable properties ranging from cluster formation in
gaseous phases,\cite{drake90,babic92,deltour97} to unexpected flows
and jets in fluid-like phases,\cite{thoroddsen01,lohse05} and to
complex organization of inhomogeneous stresses in dense fluid or solid
phases.\cite{dantu67,nagel92,claudin95} In this paper we focus on one
feature of the solid phase that has received much attention: the
distribution $P(f)$ of contact forces between grains in a system that
is supporting a macroscopic compression or shear force.  In a number
of experiments and numerical simulations involving non-cohesive
grains, $P(f)$ appears to decay approximately exponentially at large
$f$,\cite{radjai02,blair01,lovoll99,mueth98,mcnamara04} which runs
counter to the na\"{i}ve expectation of a Gaussian distribution about
some average $f$.  Several theoretical analyses of model systems have
indicated possible explanations for an exponential
tail,\cite{coppersmith96,snoeijer04a,snoeijer04b,socolar98,edwardsgm03,metzger04,metzger05}
including an analytic calculation for a special case of isostatic
packings of frictionless disks. \cite{tkachenko00} Still, a
fundamental understanding of the phenomenon has not been achieved.

To motivate the problem considered in the present work, we recall that
for generic packings of frictional and/or nonspherical hard particles
geometrical constraints permit the formation of more contacts than
would be required for supporting imposed stresses.  That is, the
contact network can contain enough contacts that the stress balance
conditions do not determine a unique configuration of the intergrain
forces. (See, for example, Ref.~\cite{unger05}.) In such cases, the
determination of $P(f)$ must involve some sort of average over the
ensemble of possible force configurations.  Edwards has suggested that
the appropriate measure in configuration space for this ensemble is a
flat one; i.e., that all possible force configurations should be
considered equally weighted.\cite{edwards} In general, such an
ensemble should include averages over different contact network
geometries as well as different stress states on a given
network.\cite{bouchaud03}

In this paper we investigate the question of whether Edwards'
hypothesis leads to exponential tails in $P(f)$ for a system of
non-cohesive grains in which the contact network forms a triangular
lattice.  We first study the case of hydrostatic compression, where
our results confirm those of Snoeijer et al,\cite{snoeijer04a,
  snoeijer04b} though we employ different boundary conditions and
different numerical methods.  We then examine the effects of diluting
the lattice to the rigidity percolation threshold.  The transition
appears to be first order, with a finite jump in the number of
available configurations at threshold.  $P(f)$ becomes substantially
broader than in the full lattice case, but still decays faster than
exponentially on the lattice sizes within our numerical reach.
Finally, we consider the effects of anisotropic imposed stresses.  We
show that strong anisotropy must produce an exponential tail in large
systems and present numerical results showing the approach to this
limit for varying degrees of anisotropy.

Because the difference between the decay produced by our model and a
true exponential decay is rather subtle, it is difficult to give firm
interpretations of most of the experimental and numerical results.
Observation of two or three decades of exponential decay could still
be consistent with the slowly developing deviation we observe in
isotropic systems.  On the other hand, a given experiment may involve
sufficiently strong anisotropies that we would expect something even
closer to a true exponential.  The recent results of Majmudar and
Behringer \cite{trush05} clearly illustrate that anisotropy due to
shearing produces distributions much closer to exponential than those
observed under isotropic compression, but we urge caution in making
direct comparisons between the model results and experiments on
disordered, deformable grains.

In the model described below, all forces are directed along the line
of contact between two grains and tensile forces are not allowed.
This may be thought of as corresponding to the case of frictionless,
non-cohesive, circular disks, though a generic set of perfectly
circular disks could not form as many contacts as are present in the
model.  Periodic boundary conditions are imposed, so there is no
distinction between bulk and boundary contacts.  By treating the force
network without regard to any distortions of the lattice, we are
considering systems of grains with elastic moduli large enough that
the boundary forces cause negligible strain; i.e., we consider either
very hard grains or very weak boundary forces.  In all cases, we fix
the contact network and study the ensemble of force configurations on
that network.  We note that fluctuations are large.  $P(f)$ in any
given force configuration may look quite different from the $P(f)$
obtained by averaging over configurations, even on the largest
lattices we have studied.

For the smallest nontrivial lattice, consisting of nine grains, we
calculate $P(f)$ both analytically and numerically, and find good
agreement.  For larger lattices, we employ numerical sampling methods
involving Monte Carlo moves that maintain the force balance
constraints at all times (in contrast to simulated annealing methods
\cite{snoeijer04a, snoeijer04b, sim_anneal}).

\section{The lattice}

\begin{figure}[tbp]
\centering
\includegraphics[clip,width=0.8\linewidth]{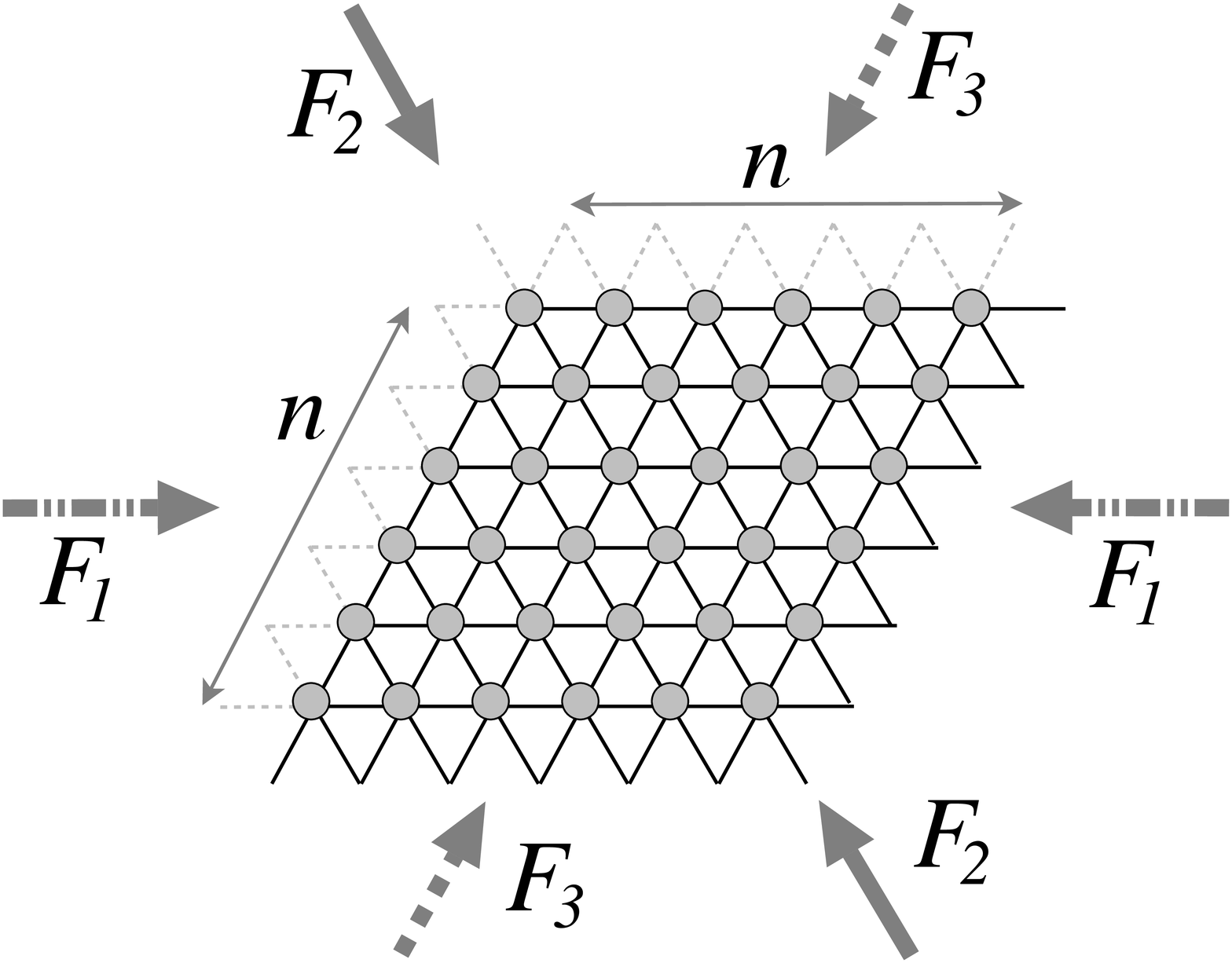}
\caption{An $n \times n$ triangular lattice (with $n=6$).  Under periodic
  boundary conditions light edges on the left are identified with
  dangling edges on the right and light edges on top are identified
  with dangling edges on the bottom.  The macroscopic stress state is
  fixed by imposing compressive forces $F_1$, $F_2$, and $F_3$ along
  each lattice direction.}
\label{fig:lattice}
\end{figure}

We study the distribution of bond strengths on an $n\times n$
triangular lattice corresponding to the contacts in a hexagonal
packing of monodisperse circular grains.  We consider a system in the
shape of a rhombus, having $n^2$ bonds or edges in each lattice
direction, and subject to periodic boundary conditions, as illustrated
in Fig.~\ref{fig:lattice}.  Each edge carries a scalar variable $f$
specifying the magnitude of the force transmitted across a contact.
In order to fix the three components of the macroscopic stress tensor
$\sigma_{ij}$ we specify the total compressive forces $F_1$, $F_2$,
and $F_3$ supported along each of the three lattice directions.

There are $3n^2$ variables in the system.  For each grain, the vector
force balance condition imposes $2$ constraints.  These constraints are
not all independent, however.  The periodic boundary conditions
guarantee that the sum of all the single-grain constraint equations is
trivially zero, leaving $2n^2 - 2$ independent constraints.  Fixing
$F_1$, $F_2$, and $F_3$ imposes 3 additional constraints, for a total
of $2n^2 + 1$ constraints.  This leaves $n^2-1$ degrees of freedom in
the force configuration.

In addition to force balance equations, there are inequality
constraints associated with the fact that the material is
non-cohesive.  No force $f$ is allowed to be negative.  It is this
condition that introduces nonlinearity in the system.  For any two
force balanced, tension free configurations on the lattice, any
weighted average with positive weights will also be force balanced and
tension free.  Sums with a negative weighting of a configuration,
however, will not always be allowed, as they may contain negative
forces on some edges.

We represent the system configuration by a vector of forces $f_i$,
$i=1\ldots 3n^2$.  The above constraints create a space of possible
configurations that fills a finite, convex $n^2-1$-dimensional volume,
with boundaries determined by the inequalities $f_i\geq0$.  Our
interpretation of the Edwards hypothesis is to assign a uniform
probability (Lebesgue) measure \cite{flatmeasure} on the space of
allowed configurations.

\section{Isotropic lattice}

In principle, $P(f)$ can be calculated in the following way.  Fix one
edge to carry force $f$; this restricts the system to a
($n^2-2$)-dimensional subset of the ($n^2-1$)-dimensional allowed
volume $V$ in configuration space.  $P(f)$ is simply the ratio of the
($n^2-2$)-dimensional ``area'' to the ($n^2-1$)-dimensional total volume. 
More formally, let $E_m$ be the set of 6 edges
touching node $m$; let $\hat{e}_g^{(m)}$, $g=1\ldots6$, be a unit vector along edge
$g$ in $E_m$ pointing towards node $m$; and let $L_k$, $k=1\ldots3$, be the set of
$n^2$ edges along lattice direction $k$.  Then we have
\begin{widetext}
\begin{equation}
P(f) = \frac{1}{V}
\int_{\lbrace f_j \geq 0 \rbrace} \left( \prod_{j=1}^{3n^2}df_j \right) \delta(f_i - f) 
\left[\prod_{m=1}^{n^2}\delta\left(\sum_{f_g \in E_m}{f_g \hat{e}_g^{(m)}}\right)\right]
\left[\prod_{k=1}^3 \delta \left(n F_k-\sum_{f_l \in L_k}{f_l}\right)\right] 
\label{eqn:P}
\end{equation}
\end{widetext}
where edge $i$ has been fixed to have value $f$.  The integral is
taken over all $f$'s.  The first delta function ensures that edge $i$
carries force $f$.  The next ensures force balance at each vertex, and
the last enforces the boundary conditions.  The integral represents
the volume of a $(n^2-2)$-dimensional slice of a polytope, therefore
$P(f)$ on an $n \times n$ lattice is a piecewise polynomial of order
$n^2-2$.

\begin{figure}[tbp]
\centering
\includegraphics[clip,width=0.8\linewidth]{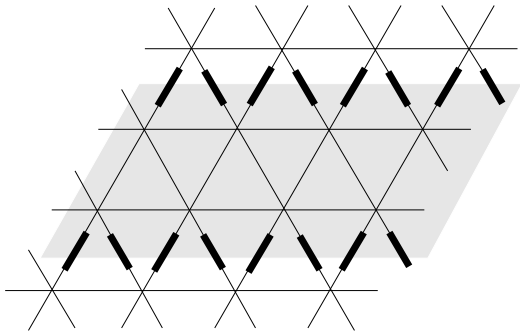}
\caption{
  A slab of material and set of edges used to prove that each layer of
  edges in $L_k$ must have a sum of forces equal to $F_k$.}
\label{fig:slab}
\end{figure}

The average force on the edges in $L_k$ is $F_k/n$.  The last factor
in Eqn.~(\ref{eqn:P}) requires comment.  For definiteness, consider
$k=2$.  Let us divide the set of edges in $L_2$ into $n$ layers, each
layer containing the $n$ edges that intersect a line along one of the
other edge directions.  The following argument shows that the sum of
the $f_j$ on each layer must be equal to $F_2$.  Fig.~\ref{fig:slab}
shows a shaded slab of the system.  The total force on this slab must
be zero, which implies that the vector sum of the forces on the edges
indicated by thick lines on one side of the slab must be equal to the
vector sum of the forces on the other side.  Since the vector sum has
a unique decomposition into contributions from the $L_2$ edges and the
$L_3$ edges, the sum of the $f_j$'s in each direction must
independently be the same on both sides of the slab.  This argument is
independent of the thickness of the slab, so in the $L_2$ direction
each of the $n$ layers must have $f_j$'s that sum to $F_2$, and
similarly for $L_1$ and $L_3$.

\begin{figure}[tbp]
\centering
\includegraphics[clip,width=0.8\linewidth]{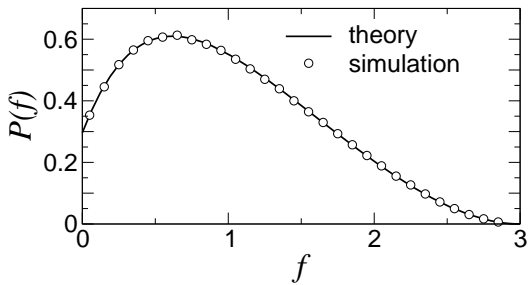}
\caption{$3\times3$ triangular lattice under isotropic stress.  The curve
  gives the force distribution, points represent numerical simulations
  of the same quantity.  $P(f)$ vanishes at $F$, the force imposed
  along each lattice direction.  Here $F=3$, fixing $\langle f \rangle = 1$.}
\label{fig:iso3x3}
\end{figure}

For the $3\times 3$ case, the integral can be evaluated analytically.
Here we state the isotropic result; the anisotropic result is
presented below.  The calculation is detailed in Appendix A.  (We have
not found an analytic expression for $P(f)$ for arbitrary lattice
size.  See Ref.~\cite{snoeijer04b}, however, for a treatment of the
case in which the sum $F_1 + F_2 + F_3$ is fixed, but not the
individual $F_k$.)  In the isotropic case ($F_1 = F_2 = F_3 = F$), we
find
\begin{eqnarray}
 & & \!\!\!\!\!P(f) = \frac{8}{45F^8}\Theta(f)\Theta(F-f)(F-f)^2  \\
 & & \!\!\!\!\!\times \, (5F^5\! +\! 73fF^4\! -\! 111f^2F^3\! +\! 125f^3F^2\! -\! 59f^4F\! +\! 9f^5) \nonumber
\label{eqn:isodist}
\end{eqnarray}
where $\Theta(f)$ is the Heaviside step function.  This expression is plotted in Fig.~\ref{fig:iso3x3}

On larger lattices we employ numerical methods to measure $P(f)$.
This requires generating numerous configurations in the allowed volume
of configuration space in a manner consistent with the Edwards flat
measure.  Previous work has implemented a simulated annealing
algorithm, which generates each configuration by starting from a
random point in the space of all possible $f_i$ and relaxing to some
point in the lower dimensional subset of interest.\cite{snoeijer04a,
  snoeijer04b} We employ a different technique in which, starting from
one force-balanced configuration -- a point in the allowed subset of
stress states -- new configurations are generated via moves that
always produce allowed configurations.  By identifying a set of moves
that span this compact space, reach any set of positive volume in a
finite time, and involve symmetric transition probabilities, we can be
sure that at sufficiently long times the space is being sampled with
uniform measure.\cite{flatmeasure} These moves are described below.

For the $n \times n$ lattice with $n \geq 3$, we construct a set of
$n^2$ {\it wheel moves}, one centered on each node.  The wheel move
associated with a given node acts on the nearest neighbor edges
(``spokes'') and next nearest neighbor edges (``rim'') of the node,
reducing one set of $f$'s and augmenting the other by the same amount.
\begin{figure}[tbp]
\centering
\includegraphics[clip,width=0.8\linewidth]{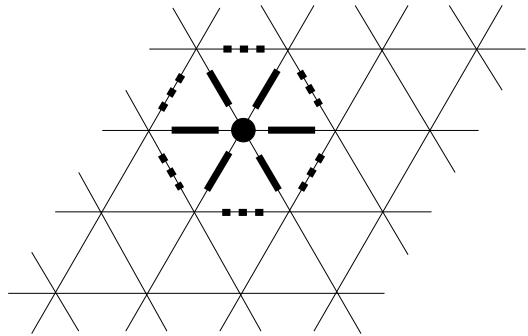}
\caption{The wheel move associated with the node marked by a disk.  Forces
  are shifted by an amount $\Delta f$, chosen to be small enough that
  no edge becomes negative.  Force on each edge marked with a thick
  solid segment is increased and on each edge marked with a thick
  dashed segment is decreased, or vice versa.}
\label{fig:wheel}
\end{figure}
The move is implemented in the following steps.  Identify the smallest
force on the spokes and call it $s_{min}$; likewise, call the smallest
force on the rim $r_{min}$.  Randomly choose a force increment $\Delta
f$ with uniform measure on the interval $[-s_{min}, r_{min}]$.  Adding
$\Delta f$ to every edge on the spokes and subtracting it from every
edge on the rim, which respects force balance on every node touching the
wheel, constitutes a wheel move.  

In the space of linear combinations of wheel moves, there is one obvious
null direction; i.e., a linear combination of wheel moves that results in 
no change to any edge of the lattice.  We prove in Appendix B that the 
number of linearly independent wheel moves is exactly 
$n^2-1$, which is the number required to span the space of force
configurations.

\begin{figure}[tbp]
\centering
\includegraphics[clip,width=0.85\linewidth]{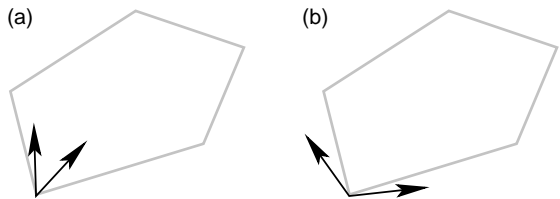}
\caption{A two-dimensional convex space, outlined in gray, with two different sets of basis vectors.  The basis in (a) is such that it is possible to move 
off the lower-lefthand corner along either basis direction, while in (b) this is not the case.}
\label{fig:stuck}
\end{figure}

Fig.~\ref{fig:stuck} illustrates a geometric subtlety that must be
taken into account when attempting to explore the allowed stress states
using wheel moves (or any other basis set of moves).
Since the given pair of basis vectors span the two-dimensional convex
space shown, it is clearly possible to navigate from any interior
point to any other by making discrete steps along the basis vectors.
It may not be possible, however, to make a move in either basis
direction if the initial configuration lies exactly on a corner.  In
higher dimensions it is likewise possible to be stuck on a corner or a
boundary of higher dimension.  It is therefore important to find
initial configurations that lie in the interior of the
space, rather than on a boundary, for the purposes of our Monte Carlo sampling technique.  An
algorithm such as the one described above, beginning at an interior
point, can come arbitrarily close to all boundaries but will never
reach them.  Regions near corners are visited infrequently but for
long times in such a way that, for sufficiently long runtime, the
space is sampled with uniform measure. \cite{flatmeasure}

Fig. \ref{fig:iso3x3} shows the agreement between the exact calculation and
simulation on the $3\times3$ lattice.  The forces are normalized by
choosing $F$ such that the average force $\langle f \rangle$ is unity.
The peak near $\langle f \rangle$ is typical for larger lattices as
well, but finite size effects are clearly evident in the tail, since
$P(f)$ must go to zero at $f=F$ (here $F=3$).
 
\begin{figure}[tbp]
\centering
\includegraphics[clip,width=1.0\linewidth]{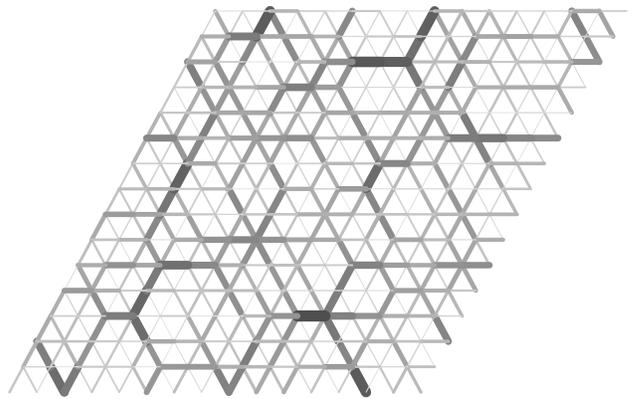}
\caption{A typical $15 \times 15$ lattice.  The force on an edge is redundantly mapped to color and width (stronger forces are darker and thicker).}
\label{fig:isolatt}
\end{figure}

\begin{figure}[tbp]
\centering
\includegraphics[clip,width=0.8\linewidth]{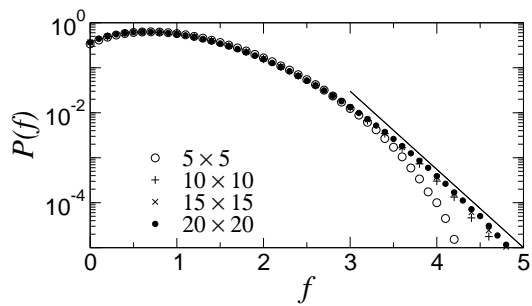}
\caption{
  Force distributions for $n \times n$ lattices under isotropic stress
  for $n=5$,$10$,$15$, and $20$.  For all cases the decay is faster
  than exponential.  The $n=15$ and $n=20$ cases are nearly identical,
  indicating convergence to a universal curve for large systems.  The
  straight line is drawn as a guide to the eye.  For each curve, $\langle f \rangle = 1$.}
\label{fig:nxn}
\end{figure}

A typical configuration for the $15\times15$ lattice is shown in Fig.~\ref{fig:isolatt}.
$P(f)$'s for the cases $n=5$, $10$, $15$, and $20$ are shown in
Fig.~\ref{fig:nxn}.  
The peak near $\langle f \rangle$ is again apparent, and for small $f$
the curves appear to coincide, though small differences exist that are
masked by the logarithmic scale.  For $f \gtrsim 3 \langle f \rangle$
the curves separate, with the $n=5$ distribution decaying most
rapidly.  The four distributions in the figure all decay faster than exponentially, and they broaden slightly with increasing $n$.

\begin{figure}[tbp]
\centering
\includegraphics[clip,width=0.8\linewidth]{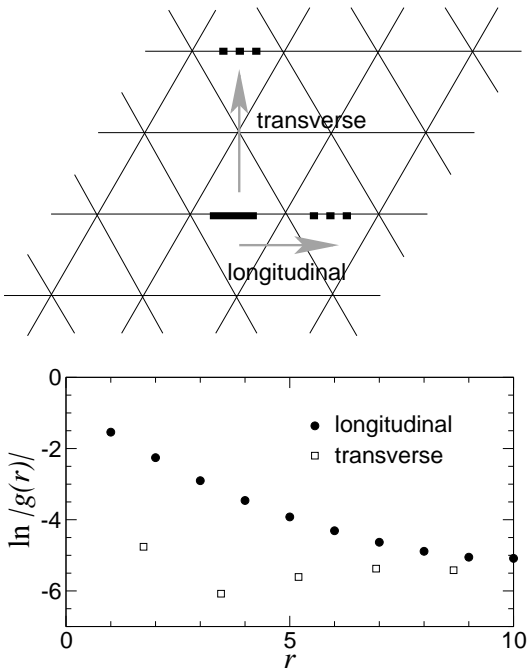}
\vspace{0.15in}

\includegraphics[clip,width=0.8\linewidth]{8b.eps}
\caption{Above, correlations are calculated in the longitudinal and transverse directions.
  Below, correlation function $g(r) = \langle f_i f_{i+r} \rangle - \langle
  f_i^2 \rangle$ on a $20\times20$ lattice.   
  The correlation is computed for edges that have the
  same orientation.  The longitudinal correlation refers to
  displacements by $r$ lattice constants along the direction of the
  edge.  The transverse correlation refers to displacements
  perpendicular to the edge.  The plots above are averaged over
  multiple reference edges.  The transverse correlations are in fact negative, but we take their absolute value to facilitate plotting on a log scale.}
\label{fig:correlation}
\end{figure}

To gain confidence that the curves are converging to a large system
limit, we measured correlations between forces as a function of the
distance between edges.  In directions both longitudinal and transverse
to a particular edge, as shown in Fig.~\ref{fig:correlation}, we see that on a $20\times20$ lattice correlations have decayed to the $1\%$ level at a distance of $10$ lattice constants.  On larger lattices we see exponential decay with a decay length of approximately $3$ lattice constants, although much more data on larger lattices would be necessary to measure this precisely.
We therefore expect to see little
difference in $P(f)$ for lattices larger than $n=20$.  Supporting this expectation, 
there is very little difference between the $n=15$ and $n=20$
distributions for $f$ up to the largest values we have measured, which
covers five decades of $P$.  Though it is conceivable that the
asymptotic form of $P(f)$ at large $f$ in our model is exponential,
the domain displayed in Fig.~\ref{fig:nxn} is the relevant one for
comparison with experiments, and it clearly shows a decay that is
faster than exponential.

\section{Diluted lattices}
Real disks are not perfectly monodisperse, so in any hexagonal packing
of hard disks some of the bonds on the triangular lattice will not
actually be present.  For a given stress state, a diluted lattice has
fewer bonds to carry the same force.  Those remaining will, on
average, carry more force, resulting in a shift in $P(f)$ toward
larger forces, which may involve a broadening of the tail.  As the
number of edges removed from the lattice increases, the force
configurations tend to become less homogeneous, with strong forces
concentrated in chain-like structures.  
\begin{figure}[tbp]
\centering
\includegraphics[clip,width=1.0\linewidth]{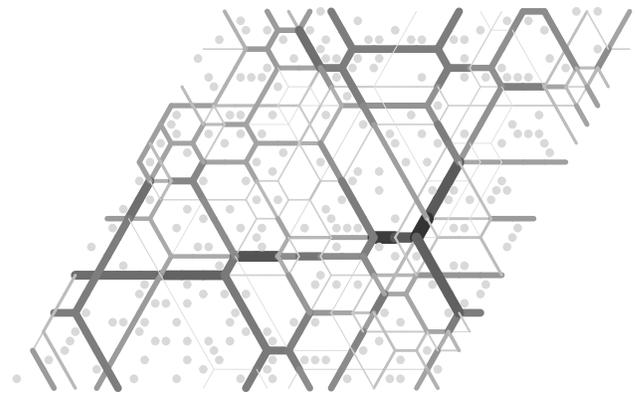}
\caption{A $15 \times 15$ lattice at rigidity percolation.  Deleted edges are covered with disks.  The force on an edge is redundantly mapped to color and width (stronger forces are darker and thicker).}
\label{fig:perlatt}
\end{figure}
As more edges are deleted there comes a point where the lattice can no
longer support the imposed stress; this is the rigidity-percolation
transition.  Fig.~\ref{fig:perlatt} shows a typical force configuration
for a randomly diluted lattice.  We study randomly diluted lattices at the transition to
see whether the broadening takes the form of an exponential decay.

The random dilution process merits further remark.  For a random
process in which each bond is removed with probability $\phi$, the
infinite triangular lattice under isotropic compression cannot support
stress for any $\phi>0$. \cite{connelly01} Thus in an $n \times n$
simulation the fraction (not the number $n_d$) of deleted edges at the
rigidity percolation threshold goes to zero as $n \rightarrow \infty$.
As an arbitrarily large real hexagonal packing does support
compression, this means that the process of bond breaking in the real
packing does not happen randomly but rather in a correlated way.
Nevertheless, our simulations employ random dilution on finite
lattices.  We also neglect for present purposes the possible buckling
instability of a force chain.  Disallowing configurations that might
buckle could only result in ensembles with fewer dilutions and
therefore would not change the conclusions described below.

We construct lattices at the rigidity percolation threshold in the
following way.  Given the imposed macroscopic stress and beginning
with an empty lattice, edges are selected at random to be added to the
lattice.  After each addition, we check to see if the lattice is
capable of supporting the imposed stress using the simplex
algorithm.\cite{simplex} We take the lattice to be at threshold when
it first supports the imposed stress, at which point we know that
there exists at least one edge such that, when removed, the system
could no longer support the stress.  The edges that have {\it not}
been added to the lattice at this stage are called ``deleted edges.''
Our construction yields a lattice in which deletions are randomly
distributed in space.  Note that the process will in general yield
some edges that are present but not connected to the edge network in a
way that allows them to bear any force.  We refer to these edges as
``effectively deleted.''  

These effectively deleted edges on a lattice supporting compressive
forces can be identified in the following way.  Consider the $6$ edges
that meet at a given node.  An edge becomes effectively deleted if its
opposite edge has been deleted (possibly effectively) and at least one
of the edges making a $120^\circ$ angle with it has also been deleted
(possibly effectively).  Under these conditions any force on the edge
in question can not be balanced by positive forces on the other edges
sharing the node.  To find all of the effectively deleted edges we
examine all nodes repeatedly until no new deletions are found.

The wheel moves used for investigation of the undiluted lattice no
longer work on the diluted lattice.  The wheel moves add or subtract a
quantity $\Delta f$ to each edge they touch, but this is impossible if
one of the edges is deleted.  Therefore each deleted edge renders
unusable the four wheel moves to which it belongs.  To salvage a set
of moves that span the appropriate space, linear combinations of the
wheel moves can be formed that leave the deleted edge unaffected.  For
a single deleted edge there are three linearly independent
combinations of the four wheel moves that edge touches which preserve
the deleted edge.  The deletion has reduced the available degrees of
freedom by one.  When many edges have been deleted, more complicated
linear combinations of wheel moves can be found that preserve the
vanishing force on all deleted and effectively deleted edges.

We refer to a particular linear combination of wheel moves that
remains a viable move on a diluted lattice as a ``multi-wheel move.''
Let the number of linearly independent multi-wheel moves be $N_m$.
Just as the undiluted lattice had $n^2$ wheel moves and $n^2-1$
degrees of freedom, the diluted lattice having $N_m$ multi-wheel moves
has $N_m-1$ degrees of freedom.  Finding a complete set of multi-wheel
moves then provides not only a means to perform numerical simulation
but also a count of the degrees of freedom in the diluted system.

A linearly independent set of $N_m$ multi-wheel moves on the diluted
lattice can be constructed as follows.  For $j=1\ldots n^2$, form a
vector $\overrightarrow{w}_j \in \mathbf{R}^{n_d}$ such that the
components of $\overrightarrow{w}_j$ specify the effect of the wheel
move centered on node $j$ on the {\it deleted} edges.  That is,
$(\overrightarrow{w}_j)_i=+1$ if the $i^\mathrm{th}$ deleted edge is a
spoke of node $j$, $-1$ if the $i^\mathrm{th}$ deleted edge is a rim
of node $j$, and $0$ otherwise.  Form the $n_d \times n^2$ matrix
$W_{ij} = (\overrightarrow{w}_j)_i$.  Consider a linear combination of
wheel moves with coefficients $m_k$, $k=1\ldots n^2$.  This linear
combination has no effect on deleted edges if and only if
$W\overrightarrow{m}=0$.  Thus $N_m = \mathrm{dim}\,N(W) = n^2-
\mathrm{rank}(W)$, where $N(W)$ is the null space of $W$.

Note that we demand the wheel moves respect both real and effective
deletions.  Were we not to specify effective deletions as well, the
algorithm would in general produce additional multi-wheel moves which
were nonzero on some effectively deleted edges.  This is because the
multi-wheel moves could also act on a lattice that supports tensile as
well as compressive forces.  On such a lattice there are fewer
effective deletions; in particular, a node could be force-balanced
having only three edges all on the same side of a line through the
node.  To model a noncohesive material, then, we must include these
additional effective deletions as well.

\begin{figure}[tbp]
\centering
\includegraphics[clip,width=0.8\linewidth]{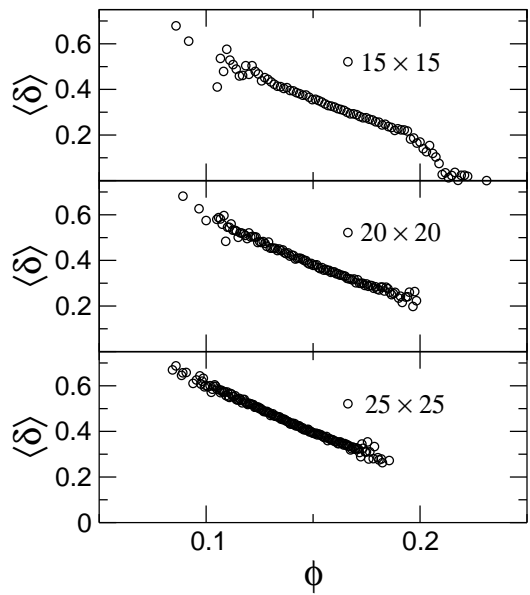}
\caption{The ratio of average number of degrees of freedom in threshold
  lattices to the number in the undiluted case $\langle \delta \rangle
  = (\langle N_m \rangle-1)/(n^2-1)$ vs. the ratio of deleted edges in
  the threshold lattice to the total number of edges $\phi =
  n_d/3n^2$.  For each plot a number of lattices at the
  rigidity-percolation threshold were generated and the degrees of
  freedom and deleted edges counted.  For large enough lattices, the
  system has a large number of degrees of freedom available as soon as
  rigidity-percolation is reached.  Note that the observed interval of
  $\phi$ values shifts as the lattice size is increased.}
\label{fig:transition}
\end{figure}

We use the following technique to investigate the degrees of freedom
in lattices at threshold.  Edges are added to a lattice one at a time
until the rigidity percolation threshold is reached, the set of wheel
moves on that threshold lattice are constructed to determine the
dimension of the volume of allowed configurations in state space
$N_m-1$, and the process is repeated for a number of threshold
lattices.  We refer to the system as having $N_m-1$ degrees of
freedom, as this is the number of multi-wheel moves it possesses.
These lattices have differing numbers of deleted edges.  For lattice
with the {\em same} number of deleted edges $n_d$ we calculate the
average ratio of the number of degrees of freedom of the dilute system
to the number in an undiluted one: $\langle \delta \rangle = (\langle
N_m \rangle - 1)/(n^2-1)$.  Figure \ref{fig:transition} shows $\langle
\delta \rangle$ plotted against the fraction of of deleted edges $\phi
= n_d/3n^2$.  A discrete jump in $\langle\delta\rangle$ emerges as
$\phi$ is decreased for systems larger than $15 \times 15$, indicating
that the transition is first order.  This suggests that $P(f)$ near
the transition point is unlikely to show qualitatively different
behavior from that of the undiluted lattice, though a quantitative
broadening of the force distribution is expected.

We measure $P(f)$ averaged over a number of threshold $20 \times 20$
lattices.  The resulting distribution is indeed much broader than the
undiluted case, but there is still curvature in the distribution on a
linear-log plot.  Numerical studies in this regime are hampered by two
factors.  First, each multi-wheel move requires the examination of
many edges to ensure that no $f_j$ becomes negative.  Second, the
maximum size of a multi-wheel move is typically quite small because
the move touches several edges that carry small forces.  The volume
that we are attempting to sample with uniform measure is still convex
but contains many corners that are not easily accessed.  The data
obtained for a given lattice may exhibit a bias depending on the
initial configuration unless the number of moves considered is
extremely large.

\begin{figure}[tbp]
\centering
\includegraphics[clip,width=0.8\linewidth]{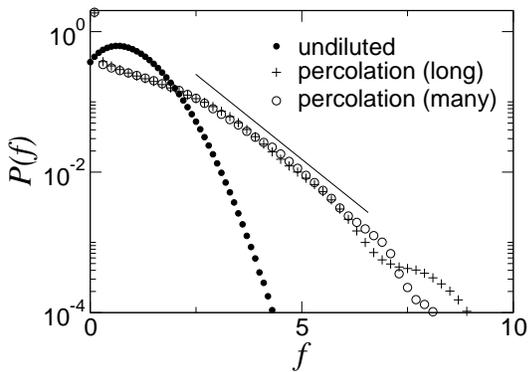}
\caption{$P(f)$ for $20\times 20$ triangular lattices diluted to the
  rigidity-percolation threshold.  The plus signs represent an average
  over $2.5\times 10^6$ moves on a single randomly diluted lattice.
  The open circles represent an average over 75 lattices, each run for
  $5\times 10^4$ moves.  The straight line is a guide to the eye.}
\label{fig:perc}
\end{figure}

We have gathered data using two procedures.  In one case we construct
a large number of threshold lattices, find a single initial
configuration for each and collect force data from $5\times 10^4$
wheel moves, then average all of the data together to determine $P(f)$
averaged over lattices.  To find initial configurations we use the
simplex method several times with different coefficients and average
the results, thereby avoiding configurations that lie on the boundary
of the allowed volume.  We cannot be sure, however, that this method
is free of systematic bias.  In the second case, we consider a single
lattice at threshold (with a typical number of degrees of freedom),
find 25 different initial configurations and run $2.5\times 10^6$
multi-wheel moves from each one, averaging the force data to obtain
the $P(f)$ associated with that particular lattice.  The results are
shown in Fig.~\ref{fig:perc}.  As may be expected given the
first-order nature of the rigidity percolation transition, neither
procedure produces an exponential tail in $P$ at large $f$.  We
conclude that random dilution to the rigidity percolation threshold
does not produce exponential tails for triangular lattices.

\section{Anisotropic lattice}

We now consider the {\it undiluted} lattice with anisotropic stresses
imposed by choosing $F_1$ different from $F_2$ and $F_3$.  For example,
one lattice direction may be subject to stronger compression than the
others, creating qualitatively distinct force distributions in the
strong and weak directions.  We will show that in strongly anisotropic
systems the strong direction contributes an exponential decay to the
tail of $P(f)$.

We consider $F_1=F+\Delta$, $F_2=F_3=F$, and parameterize the
anisotropy by $\alpha = (F+\Delta)/F$, so that $\alpha = 1$
corresponds to isotropic stress.  In terms of $\langle f \rangle$,
$F=3n\langle f \rangle/(2+\alpha)$.  The average force in the weak
direction $\langle f_w \rangle = F/n$, while that in the strong
direction is $\langle f_s \rangle = (F+\Delta)/n$.  Similarly, forces
in the weak direction can not exceed $F$, and those in the strong
direction can not exceed $F+\Delta$.  Forces in the strong direction
will populate the tail of $P(f)$.

\begin{figure}[tbp]
\centering
\includegraphics[clip,width=0.8\linewidth]{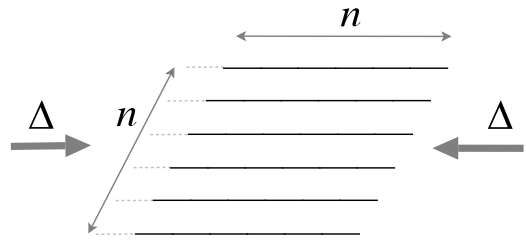}
\caption{
  $\alpha \rightarrow \infty$ limit of the $n \times n$ triangular
  lattice.  Dashed edges represent attachments under periodic boundary
  conditions.  The macroscopic stress state is fixed by imposing
  compressive force $\Delta$ along the strong lattice direction, the
  only one carrying force.  Force is transmitted along $n$ chains,
  each composed of $n$ edges, in any way such that a sum over chains
  yields $\Delta$ and all forces remain compressive.}
\label{fig:anisolimit}
\end{figure}

In the limit $\alpha \rightarrow \infty$ the system becomes much
simpler: only the strong direction carries force, doing so via $n$
chains of $n$ edges, each edge in a given chain carrying the same
force as shown in Fig.~\ref{fig:anisolimit}.  The forces on the $n$
chains must sum to $\Delta$.

Simple dimensional analysis of the scaling of the allowed volume
when one chain is fixed at force $f$ yields
\begin{equation}
\lim_{n \rightarrow \infty} P(f) \propto \lim_{n \rightarrow \infty}  \left(  1-\frac{f}{n \langle f_s \rangle}\right)^{n-2} = e^{-f/ \langle f_s \rangle}.
\end{equation}
We see that in this limit anisotropy induces an exponential tail in
$P(f)$.  We next investigate the extent to which the limiting behavior
is reflected at finite anisotropies.

To gain some intuition for the approach to the anisotropic limit we
return to the $3\times 3$ case, which can be solved analytically.  We
let $P_s(f)$ and $P_w(f)$ denote the separate distributions of forces
in the strong and weak directions, respectively. $P_w(f)$ is identical
to $P(f)$ in Eqn.~\ref{eqn:isodist}.  In the strong direction,
however, the situation is more complicated.  A complete expression is
given in Appendix C.

For the $3\times3$ case, the fully anisotropic limit $\alpha
\rightarrow \infty$ has $P(f) = \frac{2}{9}\langle f_s \rangle^{-2}(3
\langle f_s \rangle - f)$, $f \in [0,3\langle f_s \rangle$, a
piecewise linear function.
\begin{figure}[tbp]
  \centering
  \includegraphics[clip,width=0.8\linewidth]{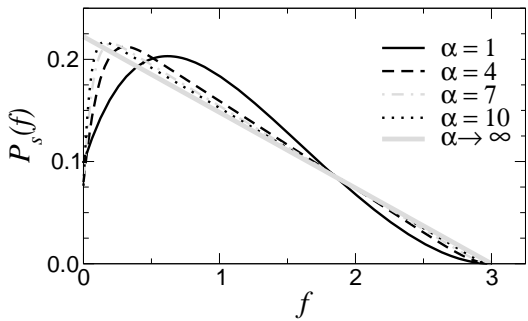}
\caption{$3\times3$ triangular lattice under anisotropic stress.  The force
  distribution $P_s(f)$ for forces in the strong direction is plotted
  for increasing anisotropy.  Each distribution is scaled to $\langle
  f_s \rangle = 1$.  The limiting form is a piecewise linear function
  of $f$.  $P_s$ for finite anisotropy has a linear region in the
  middle; this region grows in width and approaches the limiting slope
  as anisotropy is increased.}
\label{fig:3x3aniso}
\end{figure}
For $\alpha > 2$, $P(f)$ is given by $P_s^{(2b)}(f)$ (from Appendix C)
between $F$ and $\Delta$.  $P_s^{(2b)}$ is a linear function, and the
interval from $F$ to $\Delta$ grows with $\alpha$.  As $\alpha$
increases further the width of this region grows and the slope
approaches the limiting slope of $-\frac{2}{9}\langle f_s
\rangle^{-2}$.  $P_s(f)$ is plotted for increasing $\alpha$ in
Fig.~\ref{fig:3x3aniso}.

\begin{figure}[tbp]
\centering
\includegraphics[clip,width=1.0\linewidth]{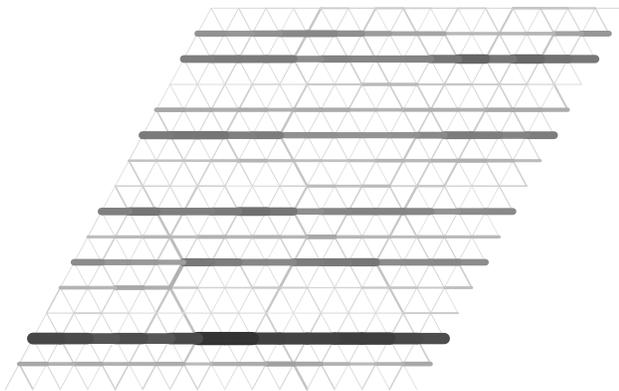}
\caption{A $15 \times 15$ anisotropic lattice with $\alpha = 5$.  The force
  on an edge is redundantly mapped to color and width (stronger forces
  are darker and thicker).  The stronger stress is imposed in the
  horizontal direction.}
\label{fig:anilatt}
\end{figure}

\begin{figure}[tbp]
\centering
\includegraphics[clip,width=0.9\linewidth]{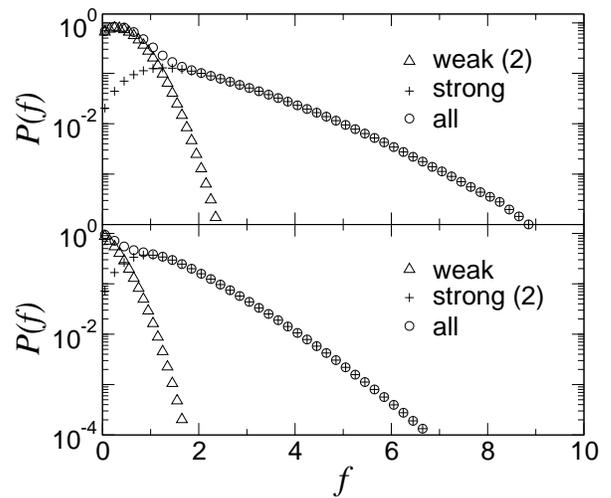}
\caption{
  $P(f)$ for strong, weak, and all lattice directions on a
  $15\times15$ lattice with $\alpha=3$.  The strong direction(s) is
  entirely responsible for the tail in $P(f)$.  Top: Two weak $F_k$'s
  and one strong.  Bottom: One weak $F_k$ and two strong.  On both plots, $\langle f \rangle = 1$.}  
\label{fig:directions}
\end{figure}

On larger lattices, numerical calculations show similar behavior to
the $3\times 3$ case, but with the limiting linear distribution
replaced by the limiting exponential of the large system limit over
the range of forces of interest.  (The distribution must be cut off at
$f = F_s$.)  Fig.~\ref{fig:anilatt} shows a typical anisotropic configuration. 
Fig.~\ref{fig:directions} shows $P(f)$ for anisotropic lattice of size
$15\times 15$, with the contributions from the weak and strong
directions shown separately.  The role of the strong direction in
broadening the distribution is clear.  Results are shown for two
cases, one in which strong forces are imposed on a single lattice
direction and another in which strong forces are imposed on two
lattice directions.  Note the difference in behavior for small $f$.
In the latter case there is no peak in $P(f)$ for small $f$.

\begin{figure}[tbp]
\centering
\includegraphics[clip,width=0.8\linewidth]{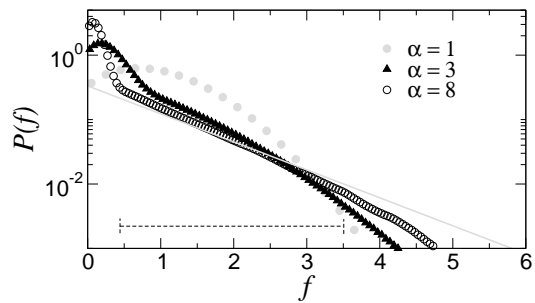}
\caption{$P(f)$ on a $15\times15$ lattice for increasing anisotropy.
  Portions of the tail for stronger anisotropies are difficult to
  distinguish from exponential decay, and the limiting behavior is a
  true exponential (gray line).  The dashed line indicates the
  interval in $f$ over which the $\alpha=8$ curve is approximately
  exponential.  $\langle f_s \rangle = 1$ for every curve.}
\label{fig:anisoexp}
\end{figure}

Fig.~\ref{fig:anisoexp} shows force distributions on a $15\times15$
lattice for several values of $\alpha$.  For sufficiently large
$\alpha$ a middle portion of $P(f)$ appears to be exponential. As
$\alpha$ increases, so does the extent of this portion.  We hold $\langle f_s \rangle$ fixed so that every curve will have the same decay length.

We conclude that $P(f)$ on the anisotropic lattice displays an
exponential tail in the following sense.  For moderate $\alpha$ and
$n$, a portion of the tail of $P(f)$ is nearly linear on a log plot.
The decay is not truly exponential because of the finite anisotropy
and finite lattice size, but the limiting behavior of $P(f)$ as $n$
and $\alpha$ are increased is a true exponential.  The region of
nearly exponential decay grows with increasing anisotropy, extending
as far as $f \approx 3\langle f_s \rangle$ for $\alpha=8$.

We note that the degree of anisotropy represented by a given value of
$\alpha$ can be compared to the anistropy supported by materials
described by an internal friction parameter.  The ratio of major and
minor principal stresses on our lattice is
$\sigma_1/\sigma_2$=$(\alpha+1)/\sqrt{3}$.  By exploiting the relation
$\sigma_1/\sigma_2 = (1+\sin{\phi})/(1-\sin{\phi})$
\cite{nedderman92}, where $\phi$ is the angle of internal friction, we
find that $\alpha = 8$ corresponds to an internal friction of
approximately $43^\circ$.

The survival of the exponential tail of strong forces in large
triangular lattices under anisotropic loading can be traced to the
discussion based on Fig.~\ref{fig:slab} concerning the sums of forces
along layers of edges in the various directions.  That argument shows
that the strong forces in one lattice direction cannot be
redistributed into the other directions due to vector force balance
constraints.  Thus for strongly anisotropic lattices, the strong
forces will follow scaling laws close to the limiting exponential form
corresponding to no force at all in the weak direction(s).  Note that
in the case of two strong directions and one weak, the two strong are
effectively independent since strong chains from the two directions
cannot interact significantly without generating a strong chain in the
supposedly weak direction and thereby reducing the degree of
anisotropy.

\section{Conclusion}
We have investigated the distribution $P(f)$ of contact forces on a
lattice of triangular bonds under the Edwards flat measure.  An
interesting question is whether one expects exponential decay in
$P(f)$.  The triangular lattice is a simple but nontrivial system for
studying this phenomenon.  

The distribution on the $n \times n$ periodic lattice decays faster
than exponentially, as reported previously.\cite{snoeijer04a} Diluting
the lattice induces significant broadening in $P(f)$, but the decay
remains faster than exponential.  Even at rigidity percolation
associated with random bond dilution of the lattice, no qualitative
change in the form of $P(f)$ is discernible.  In particular, we do not
see evidence for an exponential tail associated with the transition.
Consistent with these direct measurements of $P(f)$, we find that the
transition is first order, which generally suggests that no
qualitative changes should be expected as the percolation threshold is
approached.

On the other hand, imposing anisotropic stress on the undiluted
lattice can induce an exponential tail in $P(f)$.  In the limit of an
infinite lattice with stress imposed only along one lattice direction,
the distribution of contact forces is a pure exponential.  Numerical
simulation shows that evidence of this behavior may still be seen for
finite lattice sizes and a finite ratio of the compressive forces in
the strong and weak directions.  In such a scenario the tail of $P(f)$
is not a true exponential, but appears approximately linear on a log
plot of $P(f)$ for some interval in $f$.  For large enough
anisotropies, $P(f)$ decays three orders of magnitude from its maximum
and $f \approx 3 \langle f_s \rangle$ before deviation from
exponential decay becomes obvious.

In the triangular lattice model, anisotropy is associated with the
appearance of long force chains oriented along the strong direction, a
phenomenon that has also been observed in experiments on disordered
systems.\cite{utter04b,trush05} Extension of the numerical techniques
employed above to rigid bars that form a disordered triangulation of
the plane would be straightforward.  Wheel moves, for example, would
be defined by taking all bars sharing a given vertex to be the spokes,
though it would no longer be true that all bars receive force
increments of the same magnitude during a move.  The primary
difference between the prefect triangular lattice and disordered ones
(or even other crystalline ones) is that lattices with no sets of
collinear bars that span the system cannot support arbitrarily strong
anisotropic stresses; i.e., in most cases there would be a maximal
value of $\alpha$.  A detailed investigation of the effects of
disorder would be of interest.

\acknowledgments{We thank R.~Wolpert, B.~Rider and R.~Connelly for
  helpful conversations.  This work was supported by NSF grants
  DMR-0137119 and DMS-0244492.}

\appendix
\section{Analytic $P(f)$}
\begin{figure}[tbp]
\centering
\includegraphics[clip,width=0.8\linewidth]{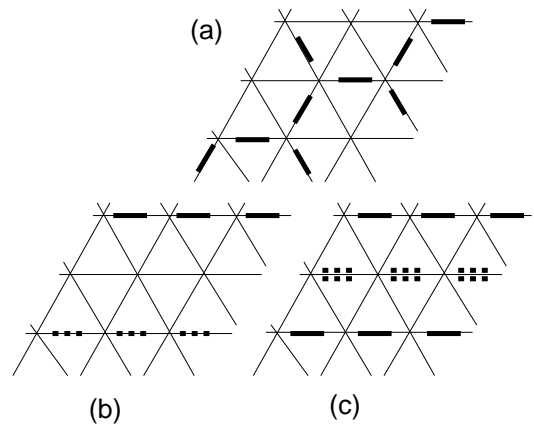}
\caption{Three of the nine basis elements employed to evaluate
  Eqn.~\ref{eqn:P} for the $3\times3$ isotropic lattice.  Solid bars
  indicate a positive contribution to the force on an edge; dashed
  bars indicate a negative contribution.  Double bars indicate a
  contribution with twice the weight.  The honeycomb in (a) is one of
  three.  The elements in (b) and (c) are each repeated for the other
  two lattice directions.}
\label{fig:basis}
\end{figure}

$P(f)$ for the $3\times3$ isotropic lattice is calculated as follows.
There are $3^2-1=8$ degrees of freedom; we take nine basis elements
subject to a constraint which is to be imposed by hand.  Three of
these elements are shown in Fig.~\ref{fig:basis}.  There are three
honeycomb elements, $\phi_i$, $i=1\ldots3$, three elements
$\psi_{j1}$, $j=1\ldots3$, and three elements $\psi_{j2}$.  The figure
depicts $\phi_1$, $\psi_{11}$, and $\psi_{12}$.  The $\psi$ elements
make no net contribution to the total force in the system.  The three
$\phi$ elements must sum to $F$; this is the additional constraint we
impose.  All elements are isotropic by construction.

Eqn.~\ref{eqn:P} can be rewritten
\begin{eqnarray}
P(f) &=& \frac{1}{N} 
 \prod_{i=1}^3 \int_0^F d\phi_i 
 \prod_{j=1}^3\int_{\phi_0/2}^{\phi_0} d\psi_{j2} 
 \int_{-\phi_0 + \psi_{j2}}^{\phi_0 - \psi_{j2}} d\psi_{j1}
 \nonumber \\
 & \times & \delta(f-(\phi_1 - 2\psi_{12}))
  \nonumber \\
 & \times & \delta(\phi_1 + \phi_2 + \phi_3 - F)
\end{eqnarray}
where $\phi_0 = \mathrm{min}(\lbrace \phi_1, \phi_2, \phi_3 \rbrace)$.
The first $\delta$-function fixes a particular edge to support force
$f$; all edges are equivalent by symmetry.  The bounds on the $\psi$'s
assure that no edge will support tensile force.  The presence of
$\phi_0$ breaks the integral into three regions.  The regions wherein
$\phi_0=\phi_2$ and $\phi_0=\phi_3$ are identical; the region where
$\phi_0=\phi_1$ must be evaluated separately because $\phi_1$ touches
the edge fixed to carry $f$.

The evaluated expression is given in Eqn.~\ref{eqn:isodist}.

\section{Proof there are $n^2-1$ independent wheel moves}
Adapting our notation from Section IV, for each $j=1\ldots n^2$ let
$\overrightarrow{w}_j \in \mathbf{R}^{3n^2}$ be the vector that
specifies the effect of the wheel move centered at node $j$ on the
collection of {\it all} edges.  Form the $3n^2 \times n^2$ matrix
$W_{ij}=\{\overrightarrow{w}_j\}_i$.  A linear combination of wheel
moves with coefficients $m_k$, $k=1\ldots n^2$ has no effect on any
edge if and only if $W\overrightarrow{m}=0$.
\begin{figure}[tbp]
\centering
\includegraphics[clip,width=0.8\linewidth]{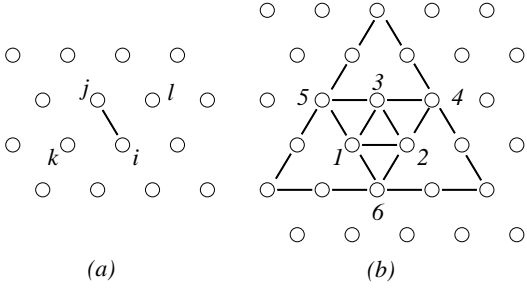}
\caption{Useful references for the proof in Appendix B.}
\label{fig:appb}
\end{figure}

{\it Lemma}: $N(W)$ is spanned by a single vector such that $m_i=m_j$
for all $i$, $j$.

{\it Remark}: Since W is $3n^2 \times n^2$, the system of equations
$Wm=0$ is highly overdetermined.

{\it Proof}: Given any edge, let $i$ and $j$ be the two vertices at
the ends of the edge, and let $k$ and $l$ be the nearest two vertices
on the perpendicular bisector (see Fig.~\ref{fig:appb}a).  The wheel
move with coefficients $\{ m_i \}$ has no effect on the given node if
and only if
\begin{equation}
m_i + m_j - m_k - m_l = 0.
\label{eqn:b1}
\end{equation}
To interpret Eqn.~\ref{eqn:b1} geometrically, let $(\mu_i, \nu_i)$ be
the coordinates of node $i$ in Fig.~\ref{fig:appb}a, and define $X_i
\in \mathbf{R}^3$ by
\begin{equation}
X_i = (\mu_i, \nu_i, m_i).
\end{equation}
Define $X_j$, $X_k$, and $X_l$ similarly.  Then Eqn.~\ref{eqn:b1} is
satisfied if and only if $X_l$ lies in the plane in $\mathbf{R}^3$
determined by $X_i$, $X_j$, and $X_k$.

Suppose the wheel move with coefficients $\{ m_i \}$ has no effect on
any edge: i.e., $W\overrightarrow{m} = 0$.  Order the nodes as
indicated in Fig.~\ref{fig:appb}b.  For the moment we regard the first
three coefficients $m_1$, $m_2$, $m_3$ as arbitrary.  Applying
Eqn.~\ref{eqn:b1} to nodes 4, 5, and 6, we deduce that $X_4$, $X_5$,
and $X_6$ all lie in the plane of $X_1$, $X_2$, and $X_3$.

With further application of Eqn.~\ref{eqn:b1} we can extend the
conclusion to all vertices of the outer triangle, and this may be
continued to conclude that all points $X_i$ lie in a single plane.
Finally, periodicity requires that this plane is horizontal: i.e.,
$m_i = m_j$ for all $i$ and $j$.  QED

\section{Anisotropic $P(f)$ on the $3\times 3$ lattice}
The calculation of $P(f)$ for the anisotropic $3\times 3$ case
proceeds similarly to the isotropic calculation.  The anisotropy
requires the division of the relevant integral according to whether
the force $f$ is greater than or less than the difference $\Delta$
between the strong $F_i$ and the weak ones. For the case $\Delta<F$,
we find
\begin{equation}
 P_s(f) = \frac{1}{N}\left\{
  \begin{array}{cc}
  P_s^{(1)}(f) & f<\Delta \\
  P_s^{(2a)}(f) & \Delta<f<F \\
  P_s^{(3)}(f) & F < f < F + \Delta,
  \end{array}
  \right.
\end{equation}
where $V$ is a normalization constant and the $P_s$'s are given below.
For the case $\Delta>F$
\begin{equation}
 P_s(f) = \frac{1}{N}\left\{
  \begin{array}{cc}
  P_s^{(1)}(f) & f<F \\
  P_s^{(2b)}(f) & F<f<\Delta \\
  P_s^{(3)}(f) & \Delta < f < F + \Delta.
  \end{array}
  \right.
  \label{eqn:anisodist}
\end{equation}

The functions $P_s$ in Eqn.~\ref{eqn:anisodist} are
\begin{widetext}
\begin{eqnarray}
P_s^{(1)}(f) & = & 
     (7\alpha-2)F^7 + 21(4\alpha-1)F^6f - 42(5\alpha+1)F^5f^2 
 + 140(2\alpha+1)F^4f^3 \nonumber \\
 & & \quad\quad\quad\quad\quad\quad\quad\quad\quad\quad\quad\quad\quad\quad 
 - 210(\alpha+1)F^3f^4 + 84(\alpha+2)F^2f^5 
 - 14(\alpha+5)Ff^6 + 12f^7; \\
P_s^{(2a)}(f) & = &
     (3\alpha^7 - 14\alpha^6 + 21\alpha^5 - 35\alpha^3 + 42\alpha^2 - 14\alpha + 2)F^7 
 + 21(\alpha^4 - 4\alpha^3 + 5\alpha^2 - 5)\alpha^2F^6f \nonumber \\
 & & \quad 
 + 21(3\alpha^4 - 10\alpha^3 + 10\alpha^2 - 15)\alpha F^5f^2 
 - 35(3\alpha^4 - 8\alpha^3 + 6\alpha^2 - 8\alpha - 5)F^4f^3 \nonumber \\
 & & \quad 
 + 105(\alpha^3 - 2\alpha^2 - \alpha - 2)F^3f^4 
 - 21(3\alpha^2 - 8\alpha - 7)F^2f^5 
 + 7(\alpha - 12)Ff^6 + 9f^7; \\
P_s^{(2b)}(f) & = & (21\alpha-4)F^7 - 21F^6f; \\
P_s^{(3)}(f) & = &   (\alpha F - f)^2
     \left[(-3\alpha^5 - 14\alpha^4 - 21\alpha^3 + 35\alpha - 42)F^5 
  + (15\alpha^4 - 56\alpha^3 + 63\alpha^2 - 35)F^4f\right. \nonumber \\
 & & \quad\quad\quad\quad\quad \left.
  + (30\alpha^2 - 56\alpha + 21)F^2f^3 
  - 3(10\alpha^2 - 28\alpha + 21)\alpha F^3 f^2 
  - (15\alpha - 14)Ff^4 + 3f^5\right].
\end{eqnarray}
\end{widetext}

\bibliographystyle{apsrev}

\end{document}